\documentclass[12pt]{iopart}
\usepackage{epsfig}

\newcommand{\be}{\begin{equation}}
\newcommand{\ee}{\end{equation}}
\newcommand{\rms}[1]{\mbox{\scriptsize {#1}}}

\newcommand{\bfr}{{\bf r}}
\newcommand{\bfR}{{\bf R}}
\newcommand{\bfE}{{\bf E}}

\newcommand{\rc}[2]{{\cal R}_{#1}^{#2}} 

\newcommand{\sq}{{\sqrt{2}}}

\newcommand{\dd}{\mbox{d}}
\newcommand{\sgn}{{\rm sgn}}
\newcommand{\bra}{\langle}
\newcommand{\ket}{\rangle}

\begin{document}
\today

\title[Electronic and center of mass transitions]
      {Electronic and center of mass transitions driven by Laguerre-Gaussian beams}
\author{A Alexandrescu$^1$\footnote[1]{Author to whom any correspondence should be addressed: aalexandrescu@tehfi.pub.ro},
    E Di Fabrizio$^2$ and D Cojoc$^{1,2}$ }
\address{ $^1$CCO -- Optoelectronics Research Center and Department of Technology and Reliability,
  ``Politehnica'' University of Bucharest, RO-61071, Romania}
\address{$^2$TASC--INFM @ Elettra Sincrotrone LILIT -- Beam line Basovizza--Trieste, IT-34012, Italy}

\begin{abstract}
We derive the interaction Hamiltonian of a Laguerre-Gaussian beam with a simple atomic system, under the
assumption of a small spread of the center of mass wave function in comparison with the waist of the Laguerre-Gaussian beam.
The center of mass motion of the atomic system is taken into account. Using the properties of regular spherical
harmonics the internal and center of mass coordinates are separated without making any multipolar expansion. Then the
selection rules of the internal and of the center of mass motion
transitions follow immediately. The influence of the winding number of the Laguerre-Gaussian beams 
on the selection rules and transition probability of the center of mass motion is discussed.
\end{abstract}

\pacs{32.80.Lg, 42.50.Vk}

\section{Introduction}

The exploration of the orbital angular momentum (OAM) of the electromagnetic field
 has been impelled by the work of Allen {\it et al.} \cite{woerdman} who
 pointed out that the Laguerre-Gaussian (LG) modes carries a well defined
 amount of OAM per unit energy. Since then the OAM properties of the radiation have gained much interest, the concepts
 of OAM being applied in several fields of optics. Ref. \cite{oam-allen} contains a collection of articles devoted or in
 connection to the OAM of the radiation field. We can take advantage of the OAM of the
 electromagnetic field to study the entanglement in a $n$-dimensional Hilbert space by spontaneous parametric 
down-conversion \cite{oam-nature}.
The interaction of LG modes with atoms as point particles has been
 studied \cite{force-oam-allen} and the transfer of the light OAM to
 atoms was observed in a four-wave mixing process \cite{tabosa}. Further, by taking into account the
 center of mass (CM) motion of the atomic system, it was shown that radiation field endowing OAM
 may entangle the internal and external OAM of the atom \cite{ashok}, and
the field OAM is imparted between the internal and external motion of the atomic system \cite{babiker,vanenk}.
The aim of this work is to investigate how the selection rules and the probability of transitions of the atomic systems change under
the influence of an LG beam, taking into account th CM motion of the atomic system.

\section{The atom-radiation system}
\label{atom-radiation-system}

We consider here the simplest atomic system, formed by a positive
charge $+e$ of mass $m_n$ and a negative charge $-e$ of mass
$m_e$, both of them spinless. The Hamiltonian of the system formed by the atom and the electromagnetic field is given by
$H=H_0+H_{\rms{int}}$, where $H_0$ is the unperturbed Hamiltonian of the atom and
$H_{\rms{int}}$ is the interaction Hamiltonian of the atom with the electromagnetic field, specified ahead in the text.

The CM coordinate of the atomic system
will be given by $\bfR=(m_{\rms{e}}\bfr_{\rms{e}}+m_{\rms{n}}\bfr_{\rms{n}})/m_{\rms{t}}$, $m_{\rms{t}}=m_{\rms{e}}+m_{\rms{n}}$
being the total mass, and their relative (internal) coordinate by
$\bfr=\bfr_{\rms{e}}-\bfr_{\rms{n}}$, where $\bfr_{\rms{e}}$ and $\bfr_{\rms{n}}$ represent the
particles coordinates. Further, we assume that the atom state can
be written as product of the CM wave function and the
internal, i.e. electronic, wavefunction
$\Upsilon(\bfR,\bfr)=\Psi_R(\bfR)\Psi(\bfr)$.
The atomic system is considered trapped in a two dimensional
harmonic potential, and its normalized wavefunction written in cylindrical
coordinates has the form:
\be
\Psi_R(\bfR)=\frac{1}{2\pi}G_{N,M}\left(\frac{R_\perp}{w_R}\right)\exp\left[\rmi(KR_z+ M\Phi)\right]
\label{center-of-mass-wavefunction}
\ee
where $G_{N,M}(x)={\cal N}\,x^{|M|}L_{n^-}^{|M|}(x^2)\exp(-x^2/2)$, $L_p^l(x)$ being the
generalized Laguerre polynomial, and $w_R$ describes the spread of the CM wavefunction. ${\cal N}=\sqrt{2\:n^-!/n^+!}/w_R$
is the normalization constant, with $n^\pm=(N\pm|M|)/2$.
The quantum number $N$ gives the energy of the harmonic
oscillator $E_R=\hbar^2(N+1)/(w_R^2m_{\rms{t}})$, and $M$ gives the
amount of angular momentum carried by the CM motion
$M\hbar$. The internal wave function is given in terms of spherical
harmonics $Y_l^m$:
\be
\Psi(\bfr)=F_{n,l}(r)\:Y_l^m(\theta,\varphi),
\label{electronic-wavefunction}
\ee where $F_{n,l}(r)$ gives the radial dependence.

We consider an electromagnetic field, in the paraxial approximation, described by a LG beam of waist $w_0$, 
having no off-axis radial nodes, propagating along the $z$ axis, and
normalized according to $\int \dd^2 r \,|E({\bf r})|^2=A$. The normalization constant $A$  
is chosen in such way that it is independent of the winding number $l$.
Under the assumption that the spread of the CM wave function is smaller in comparison with the waist of the LG
beam, the electric field reads as \cite{ashok}:
\be
\bfE(\bfr,t)=\frac{\bfE_0}{\sqrt{|l|!}}\:\left(\frac{r_\perp}{w_0}\right)^{|l|}\:
\exp[\rmi(l\varphi + k z-\omega t)].
\label{electric-field}
\ee
Next, we write the electromagnetic field in terms of regular solid spherical harmonics ${\cal R}_l^m$:
\be
\bfE(\bfr,t)=(-1)^{\frac{l+|l|}{2}}\,2^{|l|}\,\sqrt{|l|!}\,\bfE_0\:
\rc{|l|}{l}(\frac{\bfr_\perp}{w_0})\:\exp[\rmi(k z-\omega t)],
\label{electric-field-approx}
\ee
 where $\bfE_0$ is the polarization vector of the electric field, 
and $\rc{l}{m}(\bfr_\perp)={\cal C}_l^m r^lY_l^m(\theta=\pi/2,\varphi)$ is the regular solid spherical harmonic \cite{gelderen},
 with ${\cal C}_l^m=[4\pi/(2l+1)/(l-m)!/(l+m)!]^{1/2}$.
 For simplicity, the symmetry axis of the CM motion is considered to be the same as the axis of the electromagnetic LG mode. In
 the subsequent calculations the electromagnetic field is treated classically.

\section{The interaction Hamiltonian}
\label{interaction-hamiltonian-section}

We now focus on the interaction Hamiltonian which plays the
central role in revealing the influence of the winding number $l$ on
system transitions. The structure of the electromagnetic field (\ref{electric-field-approx}), i.e., in terms of spherical
harmonics, is straightway imprinted to the interaction Hamiltonian using the Power-Zienau-Wooley scheme \cite{lembessis}:
 \be H_{\rms{int}}=-\int
\dd^3r\:  {\cal P}(\bfr)\cdot\bfE(\bfr,t)+\mbox{h.c.},
\label{interaction-hamiltonian-pzw} \ee
 where ${\cal P}(\bfr)$
represent the polarization density. Writing the polarization density as a closed
integral
 \be {\cal P}(\bfr)=\sum_{\alpha=n,e}e_\alpha(\bfr_\alpha-\bfR)\int_0^1\dd\lambda\:
 \delta[(\bfr-\bfR-\lambda(\bfr_\alpha-\bfR))]
\label{polatization}
\ee
and introducing it in relation (\ref{interaction-hamiltonian-pzw}), the interaction Hamiltonian is expressed as
follows \cite{babiker}:
\be
\fl H_{\rms{int}}=\frac{e}{m_{\rms{t}}}\bfr\cdot\int_0^1
\dd\lambda\left\{m_{\rms{n}}\bfE(\bfR+\lambda\frac{m_{\rms{n}}}{m_{\rms{t}}}\bfr,t) +
m_{\rms{e}}\bfE(\bfR-\lambda\frac{m_{\rms{e}}}{m_{\rms{t}}}\bfr,t) \right\}+\mbox{h.c.}
=H_{\rms{int}}^{(1)}+H_{\rms{int}}^{(2)},
\label{interaction-hamiltonian}
\ee
where $H_{\rms{int}}^{(1)}$ and $H_{\rms{int}}^{(2)}$ refers to the terms given by
$m_{\rms{n}}\bfE(\bfR+\lambda\frac{m_{\rms{n}}}{m_{\rms{t}}}\bfr,t)+\mbox{h.c.}$ and
$m_{\rms{e}}\bfE(\bfR+\lambda\frac{m_{\rms{e}}}{m_{\rms{t}}}\bfr,t)+\mbox{h.c.}$, respectively.

 Using the expression of electromagnetic field (\ref{electric-field-approx}) in terms of regular spherical harmonics, we can separate
the CM and electronic coordinates, $\bfR$ and $\bfr$, respectively, by the translation property of 
regular spherical harmonics \cite{gelderen}:
\be
\rc{l}{m}({\bf x}\pm{\bf y})=\sum_{l^\prime=0}^l\,\sum_{m^\prime=-l^\prime}^{l^\prime}
(\pm)^{l^\prime}\rc{l^\prime}{m^\prime}({\bf x}) \rc{l-l^\prime}{m-m^\prime}({\bf y}),
\label{translation_spherical_harmonics}
\ee
and the electromagnetic field will read as:
\begin{eqnarray}
\fl\bfE(\bfR+\lambda\frac{m_{\rms{n}}}{m_{\rms{t}}}\bfr)&=&(-1)^{\frac{l+|l|}{2}}\,2^{|l|}\,\sqrt{|l|!}\:\bfE_0
\,\exp[\rmi k(R_z+\lambda\frac{m_{\mbox{\tiny n}}}{m_{\mbox{\tiny t}}}z)]
 \sum_{l^\prime=0}^{|l|} \sum_{m^\prime=-l^\prime}^{l^\prime}
{\cal C}_{l^\prime}^{m^\prime}{\cal C}_{|l|-l^\prime}^{l-m^\prime}\nonumber\\
&&\times\left(\lambda\frac{m_{\rms{n}}r_\perp}{m_{\rms{t}}w_0}\right)^{l^\prime}\:\left(\frac{R_\perp}{w_0}\right)^{|l|-l^\prime}
Y_{l^\prime}^{m^\prime}(\theta=\frac{\pi}{2},\varphi) \: Y_{|l|-l^\prime}^{l-m^\prime}(\Theta=\frac{\pi}{2},\Phi)
\label{electric-field-translation-1}
\end{eqnarray}
From the properties of the spherical harmonics functions $Y_l^m(\theta,\varphi)$ we know that in the case $\theta=\pi/2$, the functions
takes on non-zero values only if $l+m$ is an even number. The indices of the spherical harmonics functions corresponding to the
internal and CM coordinates, in the r.h.s of relation (\ref{electric-field-translation-1}), are $(l^\prime,m^\prime)$ and
$(|l|-l^\prime,l-m^\prime)$, respectively. This implies that when $l^\prime+m^\prime$, also
$|l|-l^\prime+l-m^\prime$ is an even number, so the spherical functions takes
on non-zero values at the same time and the double sum does not become zero.
Introducing the explicit values of the functions $Y_l^m(\theta=\pi/2,\phi)$ into equation (\ref{electric-field-translation-1}) we get:
\begin{eqnarray}
\fl\bfE(\bfR+\lambda\frac{m_{\rms{n}}}{m_{\rms{t}}}\bfr) = 2^{|l|}\,\sqrt{|l|!}\:\bfE_0\,
\exp[\rmi k(R_z+\lambda\frac{m_{\mbox{\tiny n}}}{m_{\mbox{\tiny t}}}z)]
\sum_{l^\prime=0}^{|l|} \sum_{m^\prime=-l^\prime}^{l^\prime}
\left(\lambda\frac{m_{\rms{n}}r_\perp}{m_{\rms{t}}w_0}\right)^{l^\prime}\left(\frac{R_\perp}{w_0}\right)^{|l|-l^\prime} &&\nonumber\\
\!\!\!\!\!\!\!\!\!\!\!\!\!\!\!\!\!\!\!\!\!\!\!
\times\frac{\exp(\rmi m\varphi)\exp[\rmi(l-m^\prime)\Phi]}{(l^\prime-|m^\prime|)!!\, (l^\prime+|m^\prime|)!!\,
(|l|-l^\prime-|l-m^\prime|)!!\, (|l|-l^\prime+|l-m^\prime|)!!}\delta_{l^\prime+m^\prime, \mbox{even} } &&
\label{electric-field-translation-2}
\end{eqnarray}
where $n!!$ stands for the double factorial. Let us analyze the influence of the double factorial $(|l|-l^\prime+|l-m^\prime|)!!$
on the double sum appearing in the above relation. 
Taking into account that $|m^\prime|\leq l^\prime$ and $l^\prime+m^\prime$ must be even,
there are two possible situations:
\begin{enumerate}
\item\label{l>0} $l>0$ leads to $|l|-l^\prime+|l-m^\prime|=-l^\prime+m^\prime$ which takes on either zero or negative values.
  The properties of the double factorial tell us that $n!!=\pm \infty$ unless $n\geq 0$. The case $-l^\prime+m^\prime<0$
  is ruled out, and we are left with $m^\prime=l^\prime$.
\item\label{l<0} $l<0$ leads to $|l|-l^\prime+|l-m^\prime|=-l^\prime-m^\prime$ also taking on either zero or negative values.
Performing a similar inference to previous case, the only valid possibility is $m^\prime=-l^\prime$.
\end{enumerate}
Summarizing the conclusion of (\ref{l>0}) and (\ref{l<0}), the double sum is reduced to a simple sum where $m^\prime=\sgn(l)\,l^\prime$.
The equation (\ref{electric-field-translation-2}) becomes
\begin{eqnarray}
\fl\bfE(\bfR+\lambda\frac{m_{\rms{n}}}{m_{\rms{t}}}\bfr)&=&2^{|l|}\,\sqrt{|l|}!\:\bfE_0\,
\exp[\rmi k(R_z+\lambda\frac{m_{\mbox{\tiny n}}}{m_{\mbox{\tiny t}}}z)]
\sum_{l^\prime=0}^{|l|}
\left(\lambda\frac{m_{\rms{n}}r_\perp}{m_{\rms{t}}w_0}\right)^{l^\prime}\:\left(\frac{R_\perp}{w_0}\right)^{|l|-l^\prime} \nonumber\\
&\times&\frac{\exp[\rmi\,\sgn(l)l^\prime\varphi]\exp[\rmi\,\sgn(l)(|l|-l^\prime)\Phi]}{(2l^\prime)!!\, [2(|l|-l^\prime)]!!}
\label{electric-field-translation-3}
\end{eqnarray}
In the above relation, one may change the internal coordinate $r_\perp$ to $r\sin\theta$ 
in order to give the interaction Hamiltonian in terms of
spherical harmonics. Replacing the expression of the plane wave by its expansion into spherical
harmonics $\exp(\rmi kr \cos\theta)=\sum_{p=0}^\infty\rmi^p\sqrt{4\pi(2p+1)}j_p(kr)Y_p^0(\theta)$, where $j_\nu(x)$
is the spherical Bessel function of the first kind of order $\nu$, the integration over $\lambda$ is performed 
 and the interaction Hamiltonian $H_{\rms{int}}^{(1,2)}$ reads as:
\begin{eqnarray}
\fl H_{\rms{int}}^{(1,2)}=(\pm)\frac{2\pi^{2}ew_0}{\sqrt{3}}\left(\frac{4}{kw_0}\right)^{|l|+1} \sqrt{|l|!} 
\, \exp(\rmi kR_z)&&\nonumber\\
\times \sum_{p=0}^{\infty} \sum_{l^\prime=0}^{|l|}\sum_{\sigma=0,\pm 1}\,\epsilon_\sigma\,
\left[\frac{(2p+1)}{\Gamma(2l^\prime+2)}\right]^{\frac{1}{2}}
\,\frac{\rmi^{p+l^\prime(1+\sgn(l))}}{(1+p+l^\prime)\Gamma(p+\frac{3}{2})\Gamma(|l|-l^\prime+1)}      && \nonumber\\
\times \left(\frac{kR_\perp}{4}\right)^{|l|-l^\prime} \left(\pm kr\frac{m_{\rms{n,e}}}{2m_{\rms{t}}}\right)^{l^\prime+p+1}
\,Y_{l^\prime}^{\sgn(l)l^\prime}(\theta,\varphi) \,Y_{1}^{\sigma}(\theta,\varphi)\,Y_{p}^{0}(\theta)
\, e^{\rms{i}\,\sgn(l)(|l|-l^\prime)\Phi} \nonumber\\
\times {_1F_2}\left({\textstyle\frac{p+l^\prime+1}{2};p+\frac{3}{2},\frac{p+l^\prime+3}{2}};-(kr\frac{m_{\rms{n,e}}}{2m_{\rms{t}}})^2
\right)+\mbox{h.c.}, &&
\label{interaction-hamiltonian-2}
\end{eqnarray}
where $\bfr\cdot\bfE_0$ was replaced by $r\sqrt{4\pi/3} \sum_{\sigma=0,\pm 1}\epsilon_\sigma Y_1^{\sigma}(\theta,\varphi)$, 
$\sigma$ being associated with the polarization of the electric field, $\epsilon_{\pm 1}=\pm(E_x\pm\rmi E_y)/\sq$ 
and $\epsilon_0=E_z$. However, in the frame of paraxial approximation, the $E_z$ component of the
electromagnetic field has a small contribution to the total electric field. The latter one can be given, to a good
approximation, in terms of the transverse electric field \cite{berry}.

In relation (\ref{interaction-hamiltonian-2}), since $p$ runs from zero to infinity and the hypergeometric
 function $_1F_2$ consists of an infinitely series, it follows that all multipole transitions are excited. Nevertheless, the factor
$r^{l^\prime+p+1}$ allows us to consider the first and the second transitions, while keeping only the first term, i.e., the constant
one, of the hypergeometric function $_1F_2$: $l^\prime+p+1=1$ gives the dipole transition, $l^\prime+p+1=2$ gives the quadrupole
transition. We note that in the dipole approximation, there is no transfer of the field OAM to the electronic motion, our result
being consistent with the concluding remark of Ref. \cite{babiker}.

\section{Selection rules and transition probabilities}
\label{selection-rules-section}

The selection rules of a certain transition are obtained by the
 transition matrix element between the final and initial state
${\cal M}_{i\rightarrow f}^{(1,2)}=\bra\Upsilon_f|H_{\rms{int}}^{(1,2)}|\Upsilon_i\ket$.
 Putting together the expression of the interaction Hamiltonian (\ref{interaction-hamiltonian-2}) and of the
 atomic wave function (\ref{center-of-mass-wavefunction})-(\ref{electronic-wavefunction}), the transition matrix element is written as:
\begin{eqnarray}
\fl {\cal M}_{i\rightarrow f}^{(1,2)} = (\pm)\frac{2\pi^{2}w_0e}{\sqrt{3}}\,\sqrt{|l|!}
\left(\frac{4}{kw_0}\right)^{|l|+1}\delta(k+K_i-K_f) &&\nonumber\\
\times \sum_{L=0}^\infty\sum_{l^\prime=0}^{|l|} \sum_{\sigma=0,\pm 1} \epsilon_\sigma\,
\left[\frac{(2p+1)}{\Gamma(2l^\prime+2)}\right]^{\frac{1}{2}} 
\,\frac{\rmi^{p+l^\prime(1+\sgn(l))}}{(l^\prime+p+1)\Gamma(p+3/2)\Gamma(|l|-l^\prime+1)} &&\nonumber\\
\times\bra G_{N_f,M_f}|\left(\frac{kR_\perp}{4}\right)^{|l|-l^\prime}|G_{N_i,M_i}\ket
\:\delta_{(M_f-M_i),\sgn(l)(|l|-l^\prime)} &&\nonumber\\
\times\bra F_{n_f,l_f}|(\pm kr\frac{m_{\rms{n,e}}}{2m_{\rms{t}}})^{l^\prime+p+1}\,
{_1F_2}\left({\textstyle\frac{p+l^\prime+1}{2};p+\frac{3}{2},\frac{p+l^\prime+3}{2}};-(kr\frac{m_{\rms{n,e}}}{2m_{\rms{t}}})^2\right)
|F_{n_i,l_i}\ket &&\nonumber\\
\times\bra Y_{l_f}^{m_f}\,|\,Y_{l^\prime}^{\sgn(l)l^\prime}\,Y_{1}^{\sigma}\,Y_{p}^{0}\,|\,Y_{l_i}^{m_i}\ket\: . &&
\label{selection-rules-matrix-element}
\end{eqnarray}
In relation (\ref{selection-rules-matrix-element}) the Dirac delta function and Kronecker delta symbol express the conservation of
the axial component of the CM momentum and the conservation of CM OAM, respectively. The bracket symbols designate
the integral over the radial components $R_\perp$ and $r$, and over the internal solid angle element $\dd(\cos\theta)\dd\varphi$,
respectively. The transition probabilities are proportional to the integrals over the radial
coordinates $R_\perp$ and $r$, being directly influenced by the absolute value of the winding number $l$.

 The selection rules which governs the electronic-type transitions are given by the integral involving the 
spherical harmonics in relation (\ref{selection-rules-matrix-element}). Using the addition theorem of
angular momentum, the spherical harmonics can be coupled and the final result expressed in
terms of Clebsch-Gordan coefficients. The parity selection rules can be easily determined taking into account the
properties of spherical harmonics, the transition being allowed only when $l_f+l_i+l^\prime+p+1$ is an even number.
\begin{table}[hbt]
\begin{center}
\begin{tabular}{cccrcrrrc}
\hline\hline
 Transitions driven by field OAM &$L$ & $ l^\prime$ & $\sgn(l)$ & $\Delta l$ & \multicolumn{3}{c}{$\Delta m$} & $\Delta M$ \\
\cline{6-8}
&&&&&${\scriptstyle -1}$ & ${\scriptstyle 0}$& ${\scriptstyle +1}$ & \\
\hline\hline
CM transition & 0 & 0 & & $\pm 1$ & $-1$& 0  & $+1$ & $l$\\
\hline
CM transition & 1 & 0 &  & $0,\pm 2$ & $-1$&0  & $+1$ & $l$\\
CM and electronic quadrupole &0 & 1 & $+1$ & $0,\pm 2$ & 0 & 1  & $+2$ & $l-1$\\
CM and electronic quadrupole &0 & 1 & $-1$ & $0,\pm 2$ & $-2$ & $-1$ & 0 & $-|l|+1$ \\
\hline\hline
\end{tabular}
\end{center}
\caption{Selection rules of transitions induced by an LG beam:
$(\Delta l,\Delta m)$ and $\Delta M$ are the changes of the electronic-type quantum numbers and of the CM OAM,
respectively. The numbers under column of $\Delta m$
correspond to different polarizations $\sigma$ of the
electromagnetic field. } \label{selection-rules-table}
\end{table}
From Table \ref{selection-rules-table} one can see that the selection rules of the dipole and quadrupole
transition are unchanged, as expected. The $z$ component of the angular momentum of the entire system is conserved,
i.e., the relation $\Delta m +\Delta M=l+\sigma$ is always fulfilled \cite{vanenk}. Here, since we are in the paraxial
approximation, the polarization of the electromagnetic field determines changes only in the internal magnetic 
quantum number $m$ of the atom, but beyond paraxial approximation this is no longer valid \cite{jauregui}.
In relation (\ref{selection-rules-matrix-element}) as $l^\prime$ runs from zero to $|l|$ the OAM of the electromagnetic
field is imparted to the CM motion and to the electronic-type motion. The maximum value of the transferred angular
momentum either to the CM motion or to the electronic-type motion does not exceed $|l|$. This feature is a peculiarity due to the
radial dependence of the electromagnetic field, assumed in relation (\ref{electric-field}); however, in general, no bound on the
value of the angular momentum transferred is set \cite{babiker}. 
Although  the selection rules are not directly governed by the winding number $l$ of the LG beam, they are sensitive to the 
sign of $l$ as one can see from Table \ref{selection-rules-table} in the case of the
quadrupole transitions with $p=0$ and $l^\prime=1$.

The influence of the winding number $l$ on the transition probability can be analyzed by performing the radial 
integral over the CM coordinate $R_\perp$
in formula (\ref{selection-rules-matrix-element}). Considering the initial and final CM states of the form
 (\ref{center-of-mass-wavefunction}), the integral is given by \cite{prudnikov}:
\begin{eqnarray}
\fl\bra G_{N_f,M_f}|\left(\frac{kR_\perp}{4}\right)^{|l|-l^\prime}|G_{N_i,M_i}\ket=
\frac{w_R^2}{2}\:{\cal N}_i\: {\cal N}_f\: \left(\frac{kw_R}{4}\right)^{|l|-l^\prime} &&\nonumber\\
\!\!\!\!\!\!\!\!\!\!\!\!\!\!\!\!\!\!\!\!\!\!\!\!\!\!\!\times\int_0^\infty \dd x\: x^{(|M_i|+|M_f|+|l|-l^\prime)/2}
\: e^{-x}\: L_{n_i^{-}}^{|M_i|}(x)L_{n_f^{-}}^{|M_f|}(x) &&\nonumber\\
\!\!\!\!\!\!\!\!\!\!\!\!\!\!\!\!\!\!\!\!\!\!\!\!\!\!\!=\left(\frac{kw_R}{4}\right)^{|l|-l^\prime}
\frac{(1+|M_i|)_{n_i^-}\:(\frac{|M_f|-|M_i|-|l|+l^\prime}{2})_{n_f^-}\:\Gamma(\frac{|M_i|+|M_f|+|l|-l^\prime}{2}+1)}
{\left(n_i^-!\: n_f^-!\: n_i^+!\: n_f^+!\right)^{1/2}} &&\nonumber\\
\!\!\!\!\!\!\!\!\!\!\!\!\!\!\!\!\!\!\!\!\!\!\!\!\!\!\!\times\: {_3F_2}(-n_i^-,{\textstyle \frac{|M_i|+|M_f|+|l|-l^\prime}{2}+1},
{\textstyle \frac{|M_i|-|M_f|+|l|-l^\prime}{2}+1};{\textstyle 1+|M_i|,\frac{|M_i|-N_f+|l|-l^\prime}{2}+1};1),&&
\label{CM_radial_integration}
\end{eqnarray}
with $M_f=M_i+\sgn(l)(|l|-l^\prime)$. 
Depending on the sign of the product $l\: M_i$, the relation (\ref{CM_radial_integration}) may
be further simplified. 

Next, we will focus on the CM transitions due to the photon absorption in the dipole interaction, 
$p=l^\prime=0$, and in the quadrupole interaction, with one OAM unit transferred to the electronic motion, 
$p=0$ and $l^\prime=1$. The dependence of the CM transition probability $P_{CM}$ on $l$, $l^\prime$ and $w_R/w_0$ is the form 
\begin{eqnarray}
\fl P_{CM}\propto\left| \frac{\rmi^{l^\prime(1+\sgn(l))}}{(1+l^\prime)\Gamma(|l|-l^\prime+1)}
\left(\frac{4}{kw_0}\right)^{|l|+1} \left[\frac{|l|!}{\Gamma(2l^\prime+2)}\right]^{\frac{1}{2}}
\bra G_{N_f,M_f}|\left(\frac{kR_\perp}{4}\right)^{|l|-l^\prime}|G_{N_i,M_i}\ket\right|^2&&\nonumber\\
\label{cm_transition_probability}
\end{eqnarray}

\begin{figure}[ht]
\epsfig{file=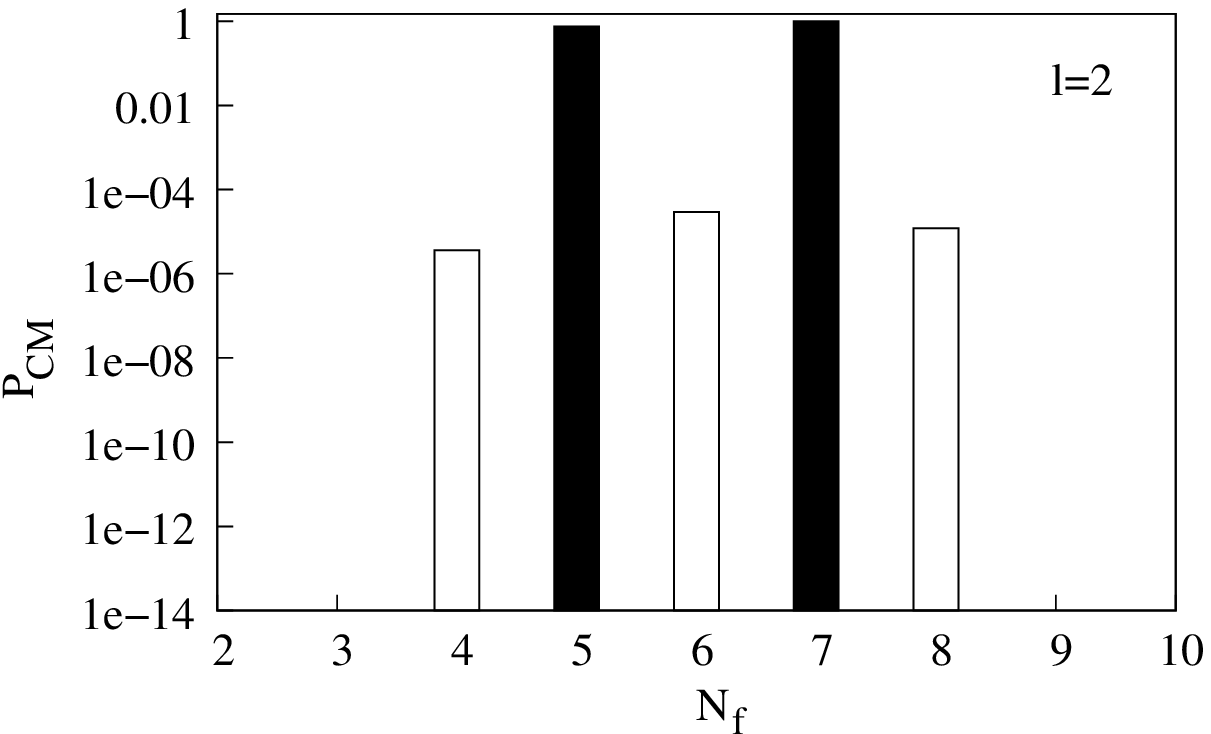, width=7cm, height=4cm}\hspace{1cm}\epsfig{file=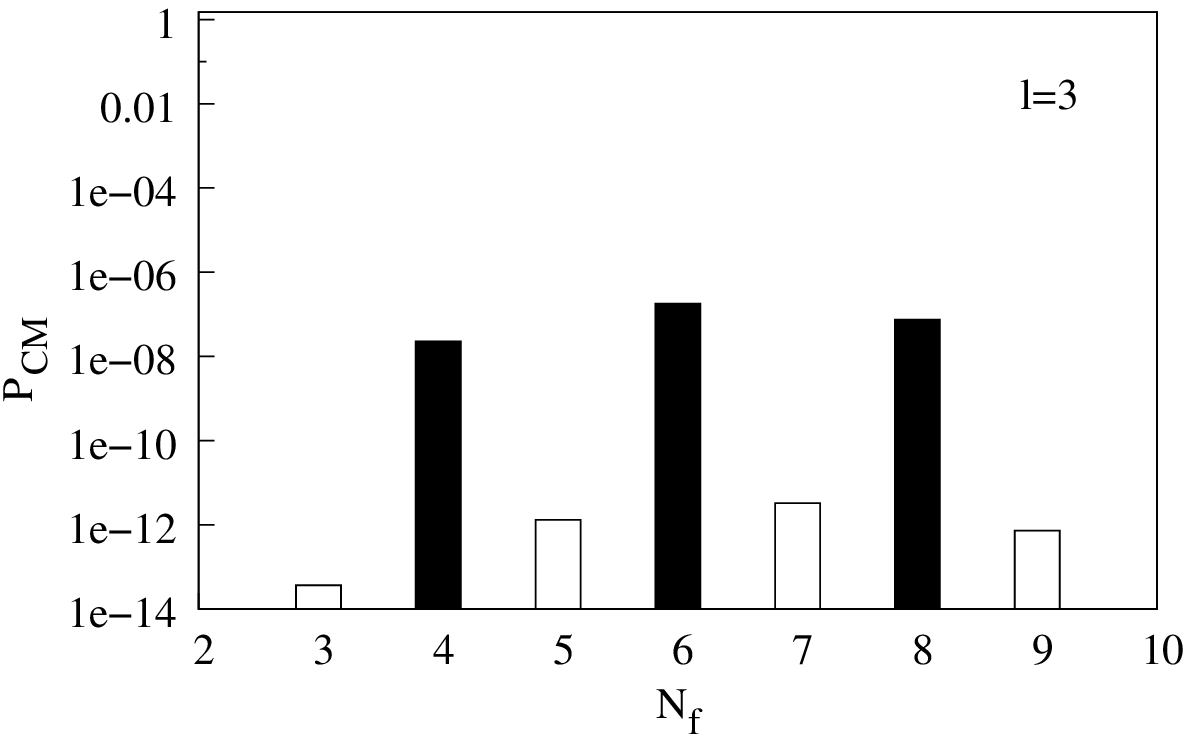, width=7cm, height=4cm}
\vspace{-0.2cm}
\begin{center}{\hspace{0.5cm}\footnotesize (a)}\hspace{7.5cm}{\footnotesize (b)}\end{center}
\caption{The CM transition probability $P_{CM}$ as function of the energy quantum number $N_f$, for $N_i=6$, $M_i=0$, 
$w_R/w_0=10^{-4}$, $l=2$ (left) and $l=3$ (right). One can see how the interaction type, i.e. dipole 
or quadrupole, changes with 
$l$ for the final states. The white and black boxes correspond to CM transitions due to the electronic dipole ($l^\prime=0$) 
and quadrupole ($l^\prime=1$) interaction, respectively. The transition probabilities are normalized to the the 
value corresponding to $N_f=6$ in the left figure.}
\label{n_dependence}
\end{figure}

The selection rules of CM transitions are given by (i) the properties of the 1D-harmonic oscillator 
(through the radial integral (\ref{CM_radial_integration})) and 
(ii) angular momentum conservation law (through the variable $\Phi$), which fixes the quantum numbers $N_f$ and $M_f$. 
In figure \ref{n_dependence} we plot the transition probability $P_{CM}$ as function of $N_f$ for $l=2$ (left) and $l=3$ (right).   
In figure \ref{n_dependence}a, the CM selection rules are $\Delta N=N_f-N_i=0,\pm 2$ 
(with the electronic transition of the dipole type, $l^\prime=0$), 
and $\Delta N=\pm 1$ (with the electronic transition of the quadrupole type, $l^\prime=1$). 
In figure \ref{n_dependence}b, for $l=3$, the number of the final CM states which can be 
addressed is larger in comparison with figure \ref{n_dependence}a, due to the multipole CM transition $|l|-l^\prime=2,3$. 

The quantum numbers of energy $N$ and of the angular momentum $M$, appearing in relation 
(\ref{center-of-mass-wavefunction}) must be of the same parity
(this comes from the fact that the wavefunction (\ref{center-of-mass-wavefunction}) describes two uncoupled 1D-harmonic 
oscillators with the quantum numbers $n_x$ and $n_y$, and from here follows $N=n_x+n_y$ and $M=n_x-n_y$). 
When one considers only the interaction in the leading orders, 
the difference in the quantum number $M_f$ will be
$M_f^{(\rms{d})}-M_f^{(\rms{q})}=\pm 1$, where (d) and (q) designates the dipole and quadrupole interaction, respectively. 
Therefore the energy number of final states reached by dipole interaction 
will have opposite parity in comparison with that one of final states reached by quadrupole interaction, see also figure 
\ref{n_dependence}.
It follows that we can not address using only one LG beam a specific state by both dipole and quadrupole interaction.
To enable both interaction types one may use superpositions of LG beams having different winding numbers $l$, e.g. by suitable
placed non-axial vortices \cite{molina}. 

From figure \ref{n_dependence} one sees that the CM transitions occurring due to the electronic dipole interaction are about
$10^{6}$ times weaker than those occurring due to the electronic quadrupole interaction. However, when one calculates the 
overall transition probability, i.e., electronic and CM, the difference between these two transition types will decrease 
to about $10^2$ times,
because the electronic quadrupole transitions are about $10^8$ times more unlikely that the electronic dipole transitions.

\begin{figure}[ht]
\begin{center}\epsfig{file=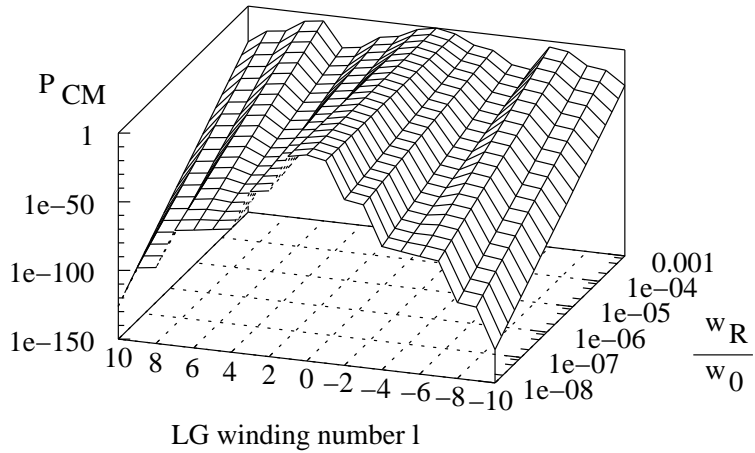, width=10cm, height=6cm}\end{center}
\caption{The CM transition probability $P_{CM}$ versus the winding number $l$ and the ratio $w_R/w_0$. 
The initial state is characterized by $N_i=6$ 
and $M_i=0$, and the final state by $N_f=12$ and $M_f=\sgn(l)(|l|-l^\prime)$. The values were normalized to $P_{CM}$ for $l=0$.}
\label{w_l_dependence}
\end{figure}

In figure \ref{w_l_dependence} we plot the CM transition probability (\ref{cm_transition_probability}) as a function of 
the winding number $l$ and of the ratio $w_R/w_0$, between the initial $N_i=6,M_i=0$ and final $N_f=12$ state.
 The transitions are driven by the dipole (quadrupole) interaction for $l$ even (odd).
 Introducing the
expression of the matrix element (\ref{CM_radial_integration}) into relation (\ref{cm_transition_probability}) the 
dependence of the transition probability $P_{CM}$ on the ratio $w_R/w_0$ will be of the form $P_{CM}\propto (w_R/w_0)^{|l|-l^\prime}$.
This behavior characterizes the approximation $w_R$ much smaller than $w_0$, 
and one may compare it with the dependence obtained from formulas
derived in Ref. \cite{babiker} in order to establish the validity range of the approximation.

\section{Conclusions}
\label{conclusions-section}

The selection rules and the matrix elements (\ref{selection-rules-matrix-element})
shows that dipole transitions affect only the CM transition. On the contrary, for quadrupole transitions there is an
exchange of one unit of angular momentum between the LG beam and the
electronic states, and no limitation exists in the exchange between the CM and
angular momentum of the beam [see first bracket in (\ref{selection-rules-matrix-element})].
In the case of an atomic system trapped in an harmonic potential the dependence of the CM transition (selection rules and
transition probability) on the LG winding number $l$ is discussed. 
There are systems \cite{chumanov} like silver
nanoparticle where the quadrupole transition are even more intense than dipolar
transitions, so that the difference of about two orders of magnitude between the electronic 
dipole and quadrupole transition can be decreased. 
In this case we suggest to perform absorption experiment using LG
beam to test the dependence of quadrupole transition on $l$. Very likely, due to
the correspondence of quadrupole transformation rules and polarizability
tensor, the dependence of CM motion on $l$, found in this paper and in Ref. \cite{babiker},
would affect Raman and Brillouin transition intensity and the relative selection rules. This
would further increase the interest in using LG beam for novel spectroscopy
studies. The application of LG beams to Raman and Brillouin transitions will be investigated
in a further work.
We conclude pointing out that even if the internal transition in quadrupole
approximation is limited to one unit of OAM, the
dependence of CM transition on $l$ is of interest in non resonant transition,
where the angular momentum of the LG beam, can be used to modulate the
transition probability.

\end{document}